# Enceladus's measured physical libration requires a global subsurface ocean

P. C. Thomas[1*], R. Tajeddine[1], M. S. Tiscareno[1,2], J. A. Burns[1,3], J. Joseph[1], T. J. Loredo[1], P. Helfenstein[1], C. Porco[4],

[1]Cornell Center for Astrophysics and Planetary Science, Cornell University, Ithaca, NY 14853 USA
[2]Carl Sagan Center for the Study of Life in the Universe, SETI Institute, 189 Bernardo Avenue, Mountain View, CA 94043
[3]College of Engineering, Cornell University, Ithaca, NY 14853 USA
[4]Space Science Institute, Boulder, CO 80304, USA

*Corresponding author E-mail:pct2@cornell.edu

**Several planetary satellites apparently have subsurface seas that are of great interest for, among other reasons, their possible habitability. The geologically diverse Saturnian satellite Enceladus vigorously vents liquid water and vapor from fractures within a south polar depression and thus must have a liquid reservoir or active melting. However, the extent and location of any subsurface liquid region is not directly observable. We use measurements of control points across the surface of Enceladus accumulated over seven years of spacecraft observations to determine the satellite's precise rotation state, finding a forced physical libration of 0.120 ± 0.014° (2σ). This value is too large to be consistent with Enceladus's core being rigidly connected to its surface, and thus implies the presence of a global ocean rather than a localized polar sea. The maintenance of a global ocean within Enceladus is problematic according to many thermal models and so may constrain satellite properties or require a surprisingly dissipative Saturn.**

## 1. Introduction

Enceladus is a 500-km-diameter satellite of mean density 1609 ± 5 kg-m$^{-3}$ orbiting Saturn every 1.4 days in a slightly eccentric (e = 0.0047) orbit (Porco et al., 2006). Much of its surface is covered by tectonic forms that have removed or modified a significant fraction of the impact crater population extant on almost all other icy moons in the outer solar system (Helfenstein et al., 2010; Bland et al., 2012). Adding to this evidence of geological activity is the discovery (Porco et al., 2006), of ongoing venting of material from fractures at high southern latitudes.

Evidence has accumulated that the jets arise from a liquid reservoir, rather than from active melting, most notably the finding that the particulates in the jets are salty, indicating freezing of droplets that likely originate in a liquid reservoir in contact with a rocky core (Waite et al., 2009; Hansen et al., 2011; Postberg et al., 2011; Porco et al., 2014; Hsu et al., 2015). Tidal heating of Enceladus driven by its elliptical orbit is the favored mechanism to form and maintain a liquid layer in such a small object that has minimal radiogenic contributions (Porco et al., 2006; Travis and Schubert 2015). The confinement of the jet activity to a ~400-m deep topographic depression poleward of ~60° S has focused attention on the possibility of a lens of liquid beneath the south polar terrain (SPT) (Collins and Goodman 2007). The stratigraphy of fractures in the south polar terrain has been interpreted as indicating long-term (perhaps on timescales >$10^6$ yr) non-synchronous rotation (Patthoff and Kattenhorn 2011) that would demand decoupling of the shell from the core and thus a global lquid layer rather than a local sea. Tracking of the Cassini spacecraft during close flybys of Enceladus yielded gravity models consistent with a mass anomaly at high southern latitudes that suggests at least a regional subsurface sea of liquid water (Iess et al., 2014). The gravity data have been reinterpreted (McKinnon 2015) as allowing for a thin, possibly discontinuous, but perhaps instead global, liquid layer.

One way to attack the problem of the liquid layer's extent is accurate measurement of the satellite's rotation (McKinnon 2015). Owing to Enceladus's slightly eccentric orbit and somewhat elongated shape (Appendix A), it is subject to periodic torques that force harmonic oscillations (called physical librations) in its orientation, on top of an overall synchronous rotation.  The magnitude of this response depends upon the object's moments of inertia and the coupling of the surface with the interior (Rambaux et al., 2011).  Precise measurements of forced libration can be accomplished by long-term stereogrammetric measurements of surface control-point networks from imaging observations, as reported for Phobos  (Oberst et al., 2014), Epimetheus (Tiscareno et al., 2009) and Mimas (Tajeddine et al., 2014).  Different techniques such as radar and laser ranging have been used to determine forced librations of  Mercury (Margot et al., 2012) and of the Moon (Rambaux and Williams, 2011)

In this paper we next review our methods of control-point calculations, and then in Section 3 we describe the basic rotational elements as related to the physical libration. Subsequently in Section 4 we report the results of the libration measurement and in Section 5 we summarize our estimation of the uncertainty of the libration measurement which is treated in detail in Appendix B.  Section 6 discusses some interior models consistent with the physical libration amplitude of Enceladus and Section 7 summarizes our results and their implications.

## 2. Methods

Control points are surface features, usually craters, whose locations are manually digitized.  Image coordinates of four or more points on crater rims are marked, and the line and sample of the center of an ellipse fit to those points (Fig. 1a) are recorded as the control-point's image coordinates.  These coordinates are then rotated with the camera's inertial orientation (C-matrix), scaled by the camera's optical parameters in combination with the relative positions of target and spacecraft, to provide body-centered (3-D) vectors.  The array of these observed (2-D) image coordinates can then be fit to predicted coordinates in the target body's coordinate frame (Davies et al., 1998). Most of the software used in this work was developed for the NEAR mission by J. Joseph (Thomas et al., 2002) with subsequent modifications by B. Carcich and J. Joseph. The processes of recording and analyzing control point data are common to most stereogrammetric measurements of planetary bodies using spacecraft imaging.

The Cassini camera's optical parameters (focal length, distortion) are sufficiently accurate that they introduce errors of well under 0.1 pixels across the detector. Geometric calibration of the ISS Narrow and Wide-Angle cameras (NAC, WAC), based on in-flight stellar images is described in Owen (2003). The NAC provides scales of 6 μrad/pixel (6 km/pixel at $10^6$ km range), and the WAC 60 μrad/pixel (60 km/pixel at $10^6$ km range). Fields of view of the two cameras are 0.35° and 3.5°.

Because achievable precision in the measurements is far better than the camera pointing information, all images require pointing corrections. In this operation, the target body's center is shifted in line and sample (X, Y). We do not generally allow the twist (rotation about the optical axis) to vary if the solution has any rotational outcome of interest. In a libration study using images obtained from high latitude, allowing the twist to vary would directly affect the inferred rotational orientation.

Images spanning all longitudes allow closure of the control network such that relative positions of all points around the object are constrained. In our solution we require at least three different measurements of a point, with minimal angular separation of 10°. Nearly all our data far exceed the minimal angular and number requirements.

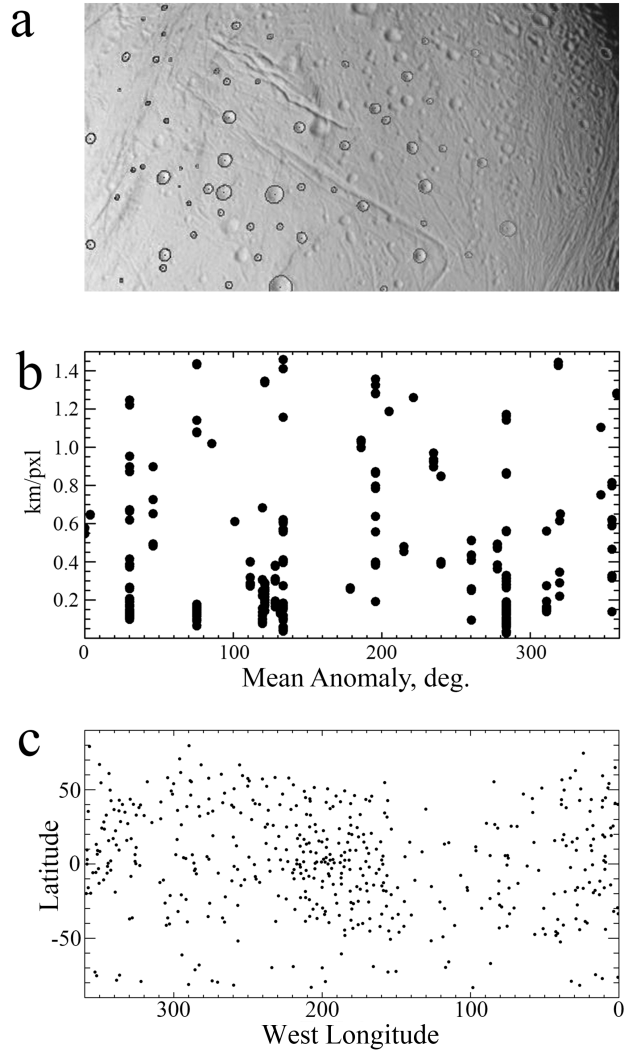

Fig. 1. Measurement of the libration amplitude. (a) Example of marked control points in a Cassini image. Image panel width 260 km. (b) The pixel scale of images used vs. mean anomaly, the angular distance from periapse. (c) Distribution of control points over the surface of Enceladus.

Because image pointing is allowed to change, the residuals in the images are determined by the relative spacing of the projections of the points in the image, rather than by total rotational offsets. Thus, for each solution, all the body-centered positions in each image are recalculated, and a change of any input data or assumed spacecraft position (including the rotation model) can affect all computed body-centered x, y, z positions. We have held the body center fixed in image coordinates for three images only, to have the coordinate origin conform to the centers found by limb-fitting (Thomas, 2010). For the overall solution, this is largely a convenience, as once the coordinate origin is reset and a best solution found with all other image pointings reset, allowing all the image centers (that is, camera-pointing) to vary did not change the average best-fit residuals (to 0.0001 pixels).

Binary kernels were developed to encode different physical libration amplitudes at increments of 0.01° over a wide range of values and with finer increments (0.001°) close to the best solution. The entire data set was then used in solutions for each assumed libration amplitude.

## 3. Rotation models

Previous observations have confirmed that Enceladus rotates synchronously to within 1.5° (Porco et al., 2014); by definition, this rate matches Enceladus's mean motion (its average angular orbital velocity) as specified by the orbit's semimajor axis via Kepler's Third Law. However, as with other bodies in the complex Saturnian system, Enceladus's semimajor axis varies by ~1 km due to perturbations from the remaining moons. The two strongest frequencies in Enceladus's semimajor axis have periods of 3.9 yr and 11.1 yr, induced, respectively, by the 2:1 corotation eccentricity near-resonance with Dione and the 2:1 Lindblad resonance with Dione (Tiscareno, 2014; Tiscareno, 2015; Giese et al., 2011a). Naturally, these temporal changes in the orbit will then cause the synchronous rotation rate to vary with time.

Some theoretical discussions (Rambaux et al., 2010) measure the orientation of Enceladus with respect to a hypothetical, long-term *constant rotation* rate, taken relative to an inertial frame and equal to the mean motion of Enceladus's average historical orbit. If we were to choose such a constant rate as the ground state for Enceladus's orientation, the temporally variable synchronous rate would oscillate about the constant rate with the periods that are dominant in the orbit's variations. Thus, in the absence of resonances (see below), any identification of the 3.9 yr and 11.1 yr periods in Enceladus's rotation state under such a scheme would merely demonstrate that dissipation within the satellite is sufficient to damp the free librations more quickly than the slowly-varying mean motion can create them, but that was never seriously in doubt (Rambaux et al., 2010; Tiscareno et al., 2009).

We use *synchronous rotation* as the ground state for Enceladus's orientation (Tiscareno et al., 2009). That is, we assume that Enceladus's rotation state remains synchronized with its mean motion, even as the mean motion changes due to the orbital resonances with Dione. We do this by keeping track of the direction of Enceladus's periapse as it precesses and thus calculating the mean anomaly $M$ (the mean angular distance from periapse) at all times. The base reference direction in our calculations is the direction to Saturn; we model the direction of Enceladus's long axis as deviating from this reference by $\psi \sin M$, where $\psi$ is a free parameter that we call the *tidal libration* amplitude (because it governs tidal torques). By Kepler's Second Law, the orbital velocity of Enceladus is higher at periapse and lower at apoapse, such that (to first order in the eccentricity $e$, which is $\ll 1$) synchronous rotation occurs if we set $\psi = 2e$, a quantity known as the *optical libration* amplitude. Thus, we represent other values of $\psi$ as the *physical libration* amplitude $\gamma = \psi - 2e$, which indicates the deviation from synchronous rotation and is the physically interesting parameter reported in this work. For Enceladus, the optical libration amplitude is $2e = 0.541°\pm 0.012°$, with the uncertainty being dominated by oscillations in the eccentricity due to the resonances with Dione (Tiscareno 2014; Tiscareno 2015). Because the libration is implemented in our kernels in terms of $\psi$, our directly measured quantity is $\psi = 0.541° + \gamma$.

This method (Tiscareno 2015) naturally causes the libration frequency to be equal to the changing mean motion at all times (that is, the libration period remains equal to the changing orbital period), such that the variations in the mean motion need not be explicitly considered; the only remaining free parameter (Rambaux et al., 2010) is the physical libration amplitude $\gamma$.

The approach described here is only valid in the absence of near-resonances between the moon's free libration period (which is several days) and any periodic terms in the variation of the moon's mean motion. If such near-resonances are absent, our approach is simpler and more easily applied. However, such an absence cannot be assumed; for example, Bills et al. (2015) showed that just such a near-resonance of the free libration period with a long-period term in the mean motion (due to the effects of tiny Hyperion, of all things) dominates the rotational librations of Titan. However, the variation of Enceladus' mean motion has no significant periodic terms other than the orbital rate of 1.37 days (the proximity of which to the free libration period is accounted for in our analysis) and the aforementioned periods of 3.9 and 11.1 years. Indeed, Rambaux et al. (2010) found that those three terms are the only ones that have any significant effect on Enceladus' rotation state.

The resulting rotation state does not have a simple Fourier transform, but it is easily calculated numerically. We encode our rotation states into a series of binary-PCK kernels in the SPICE (Spacecraft, planet, instrument, C-matrix (camera), events) navigation data system (Acton 1996), available at http://naif.jpl.nasa.gov/naif/toolkit.html. In contrast to the more common text-PCK (Planetary constants kernel) kernels, binary-PCK kernels allow for arbitrary orientation of a given body as a function of time. A binary-PCK kernel is created for each value of the libration amplitude and used as the basis for a geometric solution of the control-point network. Residuals are then calculated as described in Section 2 and statistical analysis is carried out as presented in Section 5.

## 4. Results.

We accumulated surface control-point measurements (Fig. 1a) on Enceladus between January 2005 and April 2012 employing the Cassini Imaging Science Subsystem (ISS; Porco et al., 2004). Our data are distributed throughout the orbit of Enceladus (Fig. 1b), permitting discrimination of rotational variations. They are also well spread across Enceladus's surface (Fig. 1c), providing a closed network of points and giving maximum sensitivity to changes in orientation. We located positions of 488 control points in 340 images, totaling 5873 measurements (line, sample coordinate pairs). The best solution occurs for a physical libration of $0.120 \pm 0.014°$ and has an rms residual of 0.43 pixels. The quoted uncertainty, discussed in the next section, is effectively $2\sigma$, or ~60m at the equator.

The result is not simply a value that emerges with statistical probing; the libration can be detected in single images. We obtained a solution for the control points using only images taken when Enceladus's mean anomaly was $0° \pm 40°$ or $180° \pm 40°$, portions of the orbit where the effects of any forced libration are reduced. We then compared for each of the libration amplitudes the predicted positions of the solved control points in image N1489039358 (558 m/pixel) taken at a mean anomaly of 283°, near the maximum excursion from synchronous orientation with the measured positions in the image. The comparison was made without running a solution as this action would reset all x, y, z's. In order to focus on libration effects, we calculated residuals for those points below 30° latitude. The residuals from these comparisons are shown in Fig. 2. The lowest residuals are at a physical libration angle of 0.119°. In this instance, the physical libration changes the predicted position of equatorial points by ~0.9 pixels.

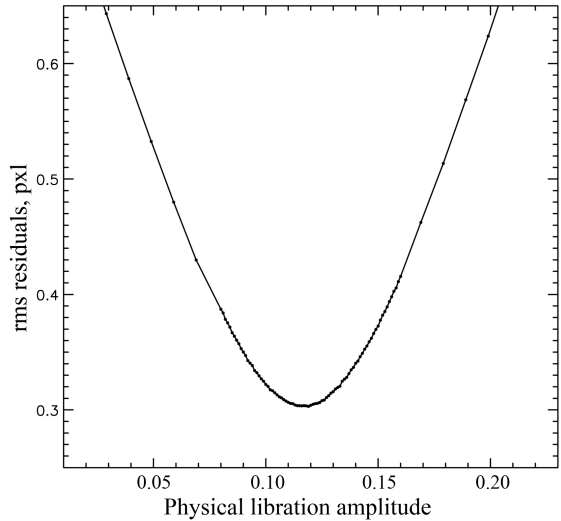

Fig. 2. Residuals of predicted positions in a single image (N1489039358) as a function of the assumed libration amplitude.

## 5. Estimation of Uncertainty

We performed a detailed analysis of the solution's expected uncertainty for the libration angle because the allowed range of forced libration angle affects interior modeling results as detailed in the following section. This uncertainty estimation was done using a Studentized residual bootstrap method described in Appendix B.

In short, we apply random errors to the best-fit results of solved image locations of each data point, the errors being scaled from the overall best-fit residuals. For each set of artificial data we obtained the best libration solution by fitting a cubic curve to the solution residuals for each of the discrete steps in libration amplitude covered by our binary kernels: 115 in total (See Fig. 2 and Figures in Appendix B). This is done 3000 times to obtain a distribution of solutions adequate to give a restrictive result equivalent to a 2-sigma uncertainty (see Appendix B). The variability of these solutions is shown in Fig. 3.

Because the libration is implemented in our kernels in terms of tidal libration, $\psi$, our directly measured best-fit value is $\psi = 0.6611° \pm 0.0078°$. However, we must account for the optical libration uncertainty which is dominated by oscillations in the eccentricity due to the resonances with Dione (Tiscareno 2014; 2015); this yields $2e = 0.541° \pm 0.012°$ (0.012° is half the full range of optical libration, and so is effectively more than 2σ). Finally, we find a derived value of the more physically interesting parameter, physical libration, $\gamma = 0.120° \pm 0.014°$.

Systematic errors can arise from camera-calibration errors, spacecraft-position errors, and biases in manually marking points on images. Calibration errors are a small part of the total residuals and further there is no expectation that they would correlate with the position in orbit when Enceladus was imaged. Spacecraft trajectory information is from reconstructed data, usually good to ~1 km in the portions of the orbit close to Saturn when images of Enceladus are obtained (Nicholson et al., 2014). Such errors, at ranges typical of the images used, would be much smaller than our computed uncertainty. Additionally, because of the oscillating nature of the libration, such errors would have to be systematically in opposite directions on the two sides of Enceladus's orbit in order to affect our solution as a function of libration amplitude. Systematic errors from digitization are not expected to correlate with orbital phase given the wide range of solar azimuths in the data throughout the images.

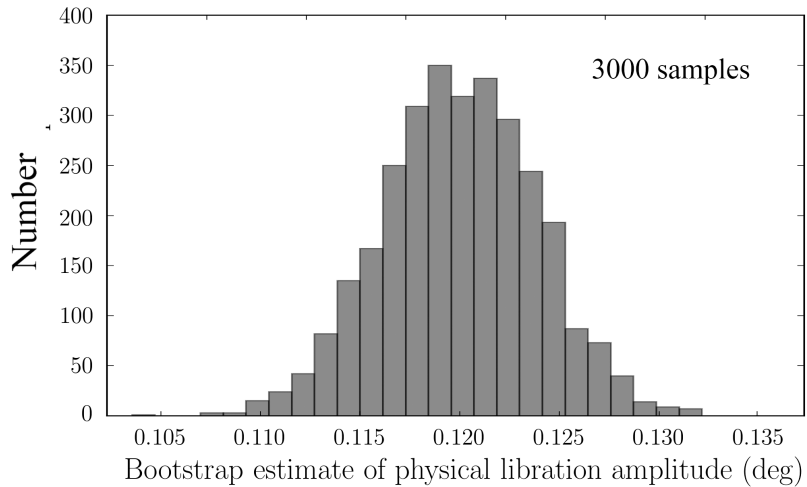

Fig. 3. Histogram showing the distribution of best-fit libration amplitudes for 3000 simulated data sets generated using the best-fit solution for the actual data (Appendix B) with randomly resampled residuals.

## 6. Interior Models

In this section we investigate several representative interior models to determine which could best explain the observed libration amplitude given above. Our knowledge of the density, moments, and libration of Enceladus of course does not allow for unique interior models, but can place severe restrictions on the internal structure of the satellite.

*6.1 Rigid models*

We first describe models for Enceladus where the core and the shell are physically coupled (i.e., without a global ocean). For a rigid body, the amplitude of the forced physical libration (Danby 1988) is

$$\gamma \approx \frac{2e}{1 - 1/(3\Sigma)}, \quad (1)$$

where we call $\Sigma = (B - A)/C$ the satellite's dynamical triaxiality, with $A$, $B$, and $C$ being the satellite's principal moments of inertia such that $A \leq B \leq C$. The numerator in Eqn. (1) is the optical libration (2e) for a synchronously rotating, spherical moon on an elliptical orbit that is amplified by resonance between the forcing (orbital) frequency and the free libration frequency, which is 3(B-A) times the orbital frequency. This yields the denominator of Eqn. (1). The moments are defined as

$$A = \int \rho(y^2 + z^2) dV$$
$$B = \int \rho(x^2 + z^2) dV \quad (2)$$
$$C = \int \rho(x^2 + y^2) dV$$

where $\rho(x, y, z)$ is the density of a volume element $dV$ at position (x, y, z) in the satellite. Here, the x, y, and z axes point along the satellite's longest (planet direction), intermediate (orbital direction), and shortest (north pole) axes.

For a homogeneous triaxial satellite, the moments of inertia, and thus the physical libration amplitude depends only on the satellite's equatorial dimensions as

$$\Sigma = \frac{a^2 - b^2}{a^2 + b^2}, \quad (3)$$

where Enceladus's triaxial half-dimensions are $a$ = 256.2 km, $b$ = 251.4 km, $c$ = 248.6 km (slightly improved from Thomas (2010) by including additional data; see Appendix A). In Eqn. (3) the numerator is proportional to the torque that Saturn exerts on Enceladus's quadrupole moment, while the denominator is proportional to the satellite's rotational inertia. These dimensions will be used for all our tested models. In this case, the free libration period is 5.77 days, far from Enceladus's 1.37 day orbital period. The libration amplitude (Table 1) for a homogeneous satellite model is $\gamma = 0.032°$, much smaller (by more than $10\sigma$) than the observed libration amplitude.

Next, we tested a two-layer model that is in hydrostatic equilibrium. In this case, the equivalent dynamical triaxiality would be

$$\Sigma = \frac{V\rho_s\left(a_s^2 - b_s^2\right) + V_c\left(\rho_c - \rho_s\right)\left(a_c^2 - b_c^2\right)}{V\rho_s\left(a_s^2 + b_s^2\right) + V_c\left(\rho_c - \rho_s\right)\left(a_c^2 + b_c^2\right)}, \tag{4}$$

where the indices "s" and "c" represent the shell and the core, respectively, $V$ is the volume, and $\rho$ is the density. We note, in Eqn. (4), that only the lead terms remain when $\rho_c = \rho_s$ (i.e., when the body is uniform) and then Eqn. (3) applies: the numerator is proportional to the sum of the torques on a homogeneous body of density $\rho_s$ plus the torque on a core of density ($\rho_c - \rho_s$) whereas the denominator is proportional to the sum of the rotational inertias for the uniform body plus the core. The core's dimensions can be expressed as functions of the polar and equatorial flattening $\alpha_c$ and $\beta_c$, respectively, as

$$a_c \approx r_c\left(1 + \frac{1}{3}\alpha_c + \frac{1}{2}\beta_c\right), \tag{5}$$

$$b_c \approx r_c\left(1 + \frac{1}{3}\alpha_c - \frac{1}{2}\beta_c\right), \tag{6}$$

where $\alpha_c$ and $\beta_c$ can be obtained by integrating the Clairaut equations (Clairaut 1743; Danby, 1988). Thus,

$$\alpha_c = \frac{12\rho_s\alpha_s + 25\rho_c q_c}{12\rho_s + 8\rho_c}, \tag{7}$$

$$\beta_c = \frac{6\rho_s\beta_s + 15\rho_c q_c}{6\rho_s + 4\rho_c}, \tag{8}$$

where $q_c = (M/m_c)(r_c/R)^3$, $M$ and $m_c$ are the masses of Saturn and the satellite's core, respectively; $r_c$ is the core's mean radius; and $R$ is the planet-satellite distance. $\alpha_s$ and $\beta_s$ are the body's polar and equatorial flattening, respectively, defined as

$$\alpha_s = \frac{(a_s + b_s)/2 - c}{(a_s + b_s)/2}, \qquad (9)$$

$$\beta_s = \frac{a_s - b_s}{b_s}; \qquad (10)$$

detailed calculations are given by Tajeddine et al. (2014). The only unknowns here are the core and shell densities, for which we chose $\rho_c$ between 2000 and 3300 kg/m³, and $\rho_s$ between 700 and 930 kg/m³. These density choices are not as arbitrary as they might seem: given Enceladus's surface composition, the upper layer's density must be close to that of water ice, including some void space, which immediately fixes the lower layer's density once the core's size is chosen so as to match the satellite's mass. The various choices of $\rho_c$ and $\rho_s$ in this model lead to predicted libration amplitudes between 0.032° and 0.034° (Table 1), still much less (by more than 10σ) than the observed libration amplitude. The slight change in the libration amplitude of this two-layer body from that for the homogeneous Enceladus given by Eqn. (1) with (3) merely reflects the small correction that the core terms make in Eqn. (4)

Because the simple solid models considered above are inconsistent with the observed libration (Table 1), we now introduce a more complex model. Gravity-mapping (Iess et al., 2014) indicates that Enceladus might have a localized South Polar Anomaly reaching up to 50°S latitude, suggesting the presence of a ~10-km thick "sea" located 30-40 km deep beneath the moon's icy crust. Such a model could also explain (Collins and Goodman 2007) a south polar depression that is observed to be about 400 meters deep (Porco et al., 2006). Therefore, we modelled a two-layer body in hydrostatic equilibrium including an axially symmetric south polar mass anomaly with a density of 1000 kg/m$^3$, an angular width of 50° (100° in total), and a thickness of 10 km. The sea is assumed to be in contact with the core, thereby accounting for the salts in Enceladus's plumes (Bland et al., 2012; Hsu et al., 2015). The moments of inertia were computed numerically by a finite-element approach in which the satellite is represented by some 10$^9$ cubes (each cube having sides of ~250 m) for the integrations of Eqns. (2). In addition, the density poleward of 50° S and to a depth of 400 m was set to 0 kg/m$^3$, in order to represent the material missing from that region. The resulting libration amplitude is the same as if the anomalies were not there (Table 1). This result occurs because all anomalies are symmetric around the z-axis, whereas the longitudinal libration depends on the torques generated by asymmetries about the equatorial axes.

*6.2 Global Ocean*

The solid interior models considered in Sec. 6.1 above are inconsistent with the observed libration amplitude of Enceladus. Accordingly, we now test a model including a global subsurface ocean that decouples the icy shell from the rocky core. We apply the approach of Van Hoolst et al. (2009), Rambaux et al. (2011) and Richard et al. (2014) to compute the libration of a satellite with a global ocean beneath its icy crust. This method accounts for the gravitational torque between the shell and the core (due to the misalignment of their principal axes of inertia) and for the ocean pressure on the shell's interior surface and the core's exterior surface. Previous studies tell us the libration amplitude is

$$\gamma_s = \frac{1}{C_c C_s} \frac{2e\left[K_s\left(K_c + 2K_{int} - n^2 C_c\right) + 2K_{int} K_c\right]}{\left(n^2 - \omega_1^2\right)\left(n^2 - \omega_2^2\right)}, \tag{11}$$

where $K_s$, $K_c$, and $K_{int}$ are the planet-shell, planet-core, and core-shell torques, respectively. Their expressions are

$$K_s = 3n^2\left[(B_s - A_s) + (B'_s - A'_s)\right], \quad (12)$$

$$K_c = 3n^2\left[(B_c - A_c) + (B'_c - A'_c)\right], \quad (13)$$

$$K_{int} = \frac{4\pi G}{5}\frac{8\pi}{15}\left[\rho_s\beta_s + (\rho_o - \rho_s)\beta_o\right]\left[(\rho_c - \rho_o)\beta_c r_c^5\right], \quad (14)$$

where the index "o" represents the ocean. Here $A'$, $B'$, and $C'$ represent the effects of the ocean pressure on the core and the shell, expressed as increments in the moments of inertia (Van Hoolst et al., 2009). $\omega_1$ and $\omega_2$ are the system's proper frequencies (free libration frequencies; for further details, see Richard et al. (2014)). These equations depend on the shapes of the different layers of the satellite, which are functions of the flattening. By following the same approach as Tajeddine et al. (2014) and integrating the Clairaut equations at the base of the ocean (assuming the core-ocean interface follows an equipotential surface), we obtain

$$\alpha_c = \frac{12(\rho_o - \rho_s)\alpha_o + 12\rho_s\alpha_s + 25\rho_c q_c}{12\rho_s + 8\rho_c}, \quad (15)$$

$$\beta_c = \frac{6(\rho_o - \rho_s)\beta_o + 6\rho_s\beta_s + 15\rho_c q_c}{6\rho_s + 4\rho_c}. \quad (16)$$

Then, by integrating the Clairaut equations from the base of the shell (assuming the ocean-shell interface follows an equipotential surface), we obtain

$$\alpha_c = \frac{12(\rho_c - \rho_o)\left(\frac{r_c}{r_o}\right)^5\alpha_c + 12\rho_s\alpha_s + 25\rho_{oc}q_o}{20(\rho_c - \rho_o)\left(\frac{r_c}{r_o}\right)^3 + 12\rho_s + 8\rho_o}, \quad (17)$$

$$\beta_c = \frac{6(\rho_c - \rho_o)\left(\frac{r_c}{r_o}\right)^5 \beta_c + 6\rho_s\beta_s + 15\rho_{oc}q_o}{10(\rho_c - \rho_o)\left(\frac{r_c}{r_o}\right)^3 + 4\rho_s + 6\rho_o}, \qquad (18)$$

where $q_o = (M/m_{oc})(r_o/R)^3$, $m_{oc} = m_o + m_c$, and $\rho_{oc}$ is the mass-weighted mean density between the core and the ocean. The polar flattenings of the core and of the ocean can be obtained by solving Eqns. (15, 17), while the equatorial flattening can be determined by solving Eqns. (16, 18).

Here, to study how an Enceladus with a global ocean might behave, we consider the densities for the core, ocean, and shell to be 2300, 1000, and 850 kg/m³, respectively, to match the moment of inertia of $0.335MR^2$ found by the gravity measurements (Iess et al., 2014). Although these densities are realistic, they could vary by as much as 10% (when including viscoelastic deformation and taking into account salts in the ocean). Applying the formulae given above, we computed Enceladus's libration amplitude as a function of the shell's thickness. We find for our interior model that an ocean 31-26 km thick, beneath an icy crust 21-26 km thick, is consistent with the observed physical libration (Table 1). The best model contains a global ocean with an icy shell as thin as ~13 km under the South Polar Terrain (SPT) (Fig. 4). Because we did not include diurnal viscoelastic deformation of the shell, which almost certainly is required by the varied structural provinces that can reduce the predicted shell thickness by up to 7 kilometers (Tajeddine et al. 2014), liquid water beneath the SPT may be even closer to the surface. We did not include possible non-uniformities in the shell and which might be driven by thermal effects and could modify the equivalent dynamical triaxiality $\Sigma$.

An unusual core shape within a solid body (McKinnon 2013, 2015; Tajeddine et al., 2014) could explain the libration amplitude. An excess in topography in the core's a-axis of 20-40 km is required to increase the libration amplitude to the observed value, but the extant gravity data do not suggest such large solid-body features. Therefore, we conclude that Enceladus's core is not rigidly connected with its surface.

Other interior models may be pursued, but regardless of changing details, the measured libration amplitude is achieved only if the shell is not connected to the core, that is, Enceladus has a global ocean. If we assume isostatic compensation, the 400-m polar depression requires a sea that is thicker, and closer to the surface under the south polar terrain than elsewhere (cf. Fig. 4).

**Table 1. Forced physical libration amplitudes for different interior models**

| Interior Model | Amplitude of forced libration |
| --- | --- |
| Homogeneous ellipsoid | 0.032° |
| 2-layer hydrostatic | 0.032° - 0.034° |
| 2-layer hydrostatic, including "polar sea" and depression | 0.032° - 0.034° |
| Ellipsoidal core, global ocean, ellipsoidal shell (23 km) (2300, 1000, 850 kg/m$^3$) | 0.120° |
| --- | --- |
| Measured Value | 0.120° ± 0.014° |

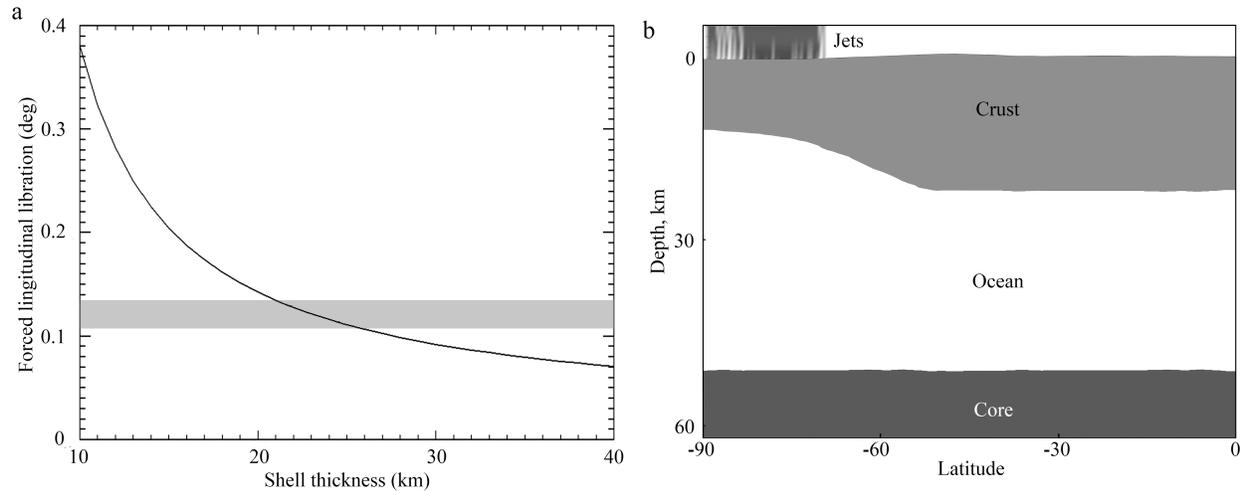

Fig. 4. Possible structure of Enceladus. (a) The amplitude of the physical forced libration as a function of the icy shell's thickness for the simplest global ocean model. The gray bar shows the range of allowed libration amplitudes at the 2σ level. (b) Schematic cross section of the upper 60 km of Enceladus from the South Pole to the equator. The surface profile displays the longitudinally averaged topography including the slight south polar depression and the latitudinal locations of jets (Porco et al., 2014). Vertical exaggeration is 4.5. Additional latitudinal and longitudinal variations in the thickness of crust and ocean are almost certain to exist. The non-uniqueness of interior models does not affect the conclusion of the existence of a global subsurface ocean.

## 7. Summary and implications

We have shown broadly that the physical libration of Enceladus, $0.120 \pm 0.014°$, is too large for there to be a rigid connection between core and crust. This finding confirms that the subsurface liquid feeding the south polar jets is part of a global subsurface ocean. The exact characteristics of this ocean, such as latitudinal and longitudinal variations in thickness and the related shell and core characteristics, are not yet uniquely defined.

Gravity data were first understood as requiring at least a regional south polar sea beneath the surface but could not discriminate a global subsurface ocean (Iess et al., 2014). A re-interpretation of the gravity data, including a higher-order rotational correction, favors a very thin, wide-spread, perhaps global ocean (McKinnon 2015).

Models of the coupled dynamical and thermal histories of Enceladus (Roberts and Nimmo 2008; Tobie et al., 2008; Meyer and Wisdom 2008; Behounkova et al., 2012) have so far been inconclusive as to whether a global ocean could be maintained over long geologic periods. Such models depend upon Enceladus's eccentricity history, its initial thermal state, the amount of tidal energy dissipated in Saturn, mechanisms for forming melted (or soft) portions that enhance tidal energy dissipation, a range of plausible physical characteristics of the shell and core, and the power currently being produced. A high dissipation (low Q) in Saturn could allow Enceladus to maintain its current heat flow (Lainey et al., 2012) and thus sustain an ocean over geologic times without requiring cyclical variations in heating from orbital interactions with other satellites (cf. Ojakangas and Stevenson 1986).

Surface geology indicates past episodes of tectonic and thermal activity somewhat similar to that in the present SPT that are approximately centered on the leading point (0°N, 90°W) and trailing point (0°N, 270°W) (Spencer et al. 2009; Crow-Willard and Papalardo 2015). These separate tectonic regions, positioned on principal moment axes, suggest a history of thermal activity and possibly of polar wander that has involved discrete epochs. Viscously relaxed impact craters (Bland et al., 2012), flexurally supported topography (Giese et al., 2008), and unstable tectonic extension of ridge-and-trough features (Bland et al., 2007) all indicate that the leading and trailing regions sustained higher thermal fluxes in the past than today. Geological features in these provinces reminiscent of active SPT features include ropy-folds or funiscular terrain on the leading region (Helfenstein et al., 2010) and topographic features that resemble extinct "tiger stripes" on both the leading and trailing regions (Spencer et al., 2009; Helfenstein et al., 2010; Crow-Willard and Pappalardo 2015).

Although surface geology indicates a past concentration of thermal and tectonic activity similar to those associated with the thin shell of today's SPT, geological evidence for the longevity of a global ocean is more difficult to establish. Possible surface geological evidence for a past global ocean includes patterns of fracture in the SPT interpreted as showing slow, long-term non-synchronous rotation (Patthoff and Kattenhorn, 2011), and the distribution of cycloidal ridges and cracks interpreted as evidence of a floating ice shell decoupled from the deep interior by a global ocean (Giese et al. 2011b). The relation of local, enhanced tectonics and heat flow to the global ocean, and the occurrence of these features near present principal axes of Enceladus, are clearly important future topics of inquiry.

**Acknowledgements**

Funded in part by the Cassini Project under JPL subcontract # 1403280. M.S.T. acknowledges funding from NASA Outer Planets Research (NNX10AP94G) and NASA Cassini Data Analysis (CDAP) (NNX13AG16G and NNX15AL21G). P.H. acknowledges CDAP Grant NNX12AG82G. We thank T. Ansty, J. Booth, B. Carcich, K. Consroe, M. Evans, M. Hedman, J. Lunine, P. Nicholson, A. Richard, D. Ruppert, T. Shannon, P. Smith. A prompt and unusually well documented reviewing process helped improve the paper.

## Appendix A. Shape of Enceladus

Limb profiles used as in Thomas (2010) to update the shape of Enceladus. Results and images used are given below.

| a, km | b, km | c, km | F | Rm | res, km | pts | images |
|---|---|---|---|---|---|---|---|
| 256.2±0.3 | 251.4±0.2 | 248.6±0.2 | 0.36 | 252.24±0.2 | 0.48 | 41780 | 54 |

F = (b-c)/(a-c), Rm = mean radius, res = average residual, pts = number of data
Images are Narrow Angle (NAC) unless noted as Wide-Angle (WAC).
In table below, lat and lon are subspacecraft coordinates at time of image.

**Images used in shape determination**

| image | #pts | rms pix | km/pixel | rms km | lat | lon | |
|---|---|---|---|---|---|---|---|
| 1484507013 | 321 | 0.196 | 3.21 | 0.63 | 3.24 | 70.14 | |
| 1484507146 | 308 | 0.194 | 3.19 | 0.62 | 3.26 | 70.33 | |
| 1484519143 | 495 | 0.236 | 2.23 | 0.53 | 4.92 | 84.25 | |
| 1484519362 | 466 | 0.235 | 2.21 | 0.52 | 4.94 | 85.44 | |
| 1484519736 | 486 | 0.276 | 2.18 | 0.60 | 5.01 | 84.77 | |
| 1484532418 | 798 | 0.361 | 1.45 | 0.52 | 7.28 | 89.97 | |
| 1484532451 | 771 | 0.374 | 1.44 | 0.54 | 7.27 | 89.96 | |
| 1484577892 | 470 | 0.305 | 1.25 | 0.38 | -2.65 | 132.39 | |
| 1484578385 | 517 | 0.342 | 1.26 | 0.43 | -2.78 | 134.07 | |
| 1487255259 | 789 | 0.333 | 1.44 | 0.48 | -0.81 | 174.70 | |
| 1487255359 | 785 | 0.369 | 1.43 | 0.53 | -0.81 | 175.41 | |
| 1487262380 | 627 | 0.297 | 1.14 | 0.34 | -0.91 | 205.15 | |
| 1487264695 | 679 | 0.312 | 1.08 | 0.34 | -0.92 | 215.32 | |
| 1487300648 | 587 | 0.388 | 1.28 | 0.50 | 2.13 | 317.30 | WAC |
| 1489029550 | 609 | 0.438 | 1.17 | 0.51 | -0.41 | 143.87 | |
| 1489029674 | 613 | 0.443 | 1.16 | 0.52 | -0.41 | 144.20 | |
| 1489029967 | 631 | 0.350 | 1.14 | 0.40 | -0.42 | 144.97 | |
| 1489034047 | 831 | 0.347 | 0.87 | 0.30 | -0.49 | 156.24 | |
| 1489039292 | 1359 | 0.521 | 0.56 | 0.29 | -0.62 | 171.98 | |
| 1489039325 | 1357 | 0.551 | 0.56 | 0.31 | -0.63 | 172.09 | |
| 1495319152 | 630 | 0.301 | 1.28 | 0.39 | -33.33 | 234.45 | |
| 1500041648 | 603 | 0.355 | 1.22 | 0.43 | -36.23 | 141.74 | |
| 1500050626 | 1059 | 0.740 | 0.67 | 0.50 | -43.01 | 166.38 | |
| 1500051528 | 1149 | 0.676 | 0.62 | 0.42 | -43.65 | 169.11 | |
| 1500068930 | 264 | 0.163 | 2.52 | 0.41 | 47.04 | 41.64 | WAC |
| 1511806152 | 995 | 0.456 | 0.90 | 0.41 | 0.94 | 137.37 | |
| 1514141354 | 1307 | 0.759 | 0.65 | 0.49 | -0.20 | 114.87 | |
| 1516160530 | 823 | 0.600 | 0.90 | 0.54 | -0.44 | 222.53 | |
| 1516160714 | 836 | 0.552 | 0.90 | 0.49 | -0.43 | 223.00 | |

```
1516170051   767  0.777  0.92  0.72   -0.09  241.16
1516171363   778  0.748  0.94  0.70   -0.04  242.83
1516171418   761  0.736  0.94  0.69   -0.04  242.90
1541672539   279  0.118  2.39  0.28  -57.99  129.92
1547630665   234  0.107  3.94  0.42  -72.44  203.64
1561727555   447  0.309  1.69  0.52   -2.31  191.03
1561728992   424  0.292  1.79  0.52   -2.18  196.86
1584051749   789  0.480  0.79  0.38  -69.16  345.43
1584052713   752  0.433  0.86  0.37  -68.99  348.56
1597183004   943  0.529  0.77  0.41  -63.45  279.65 WAC
1637475217  1233  0.657  0.59  0.39   -0.82   51.35
1637475867  1162  0.612  0.62  0.38   -0.73   52.55
1637479601  1023  0.687  0.80  0.55   -0.42   59.13
1637479929   883  0.511  0.81  0.42   -0.38   59.65
1649350305   930  0.387  1.19  0.46   -0.37  289.81
1649350551   780  0.403  1.19  0.48   -0.34  289.35
1660446193  1297  0.772  0.56  0.43    8.89   63.68
1671602206  1264  0.561  0.61  0.34    0.17  189.49
1696191074   857  0.627  0.88  0.55   -0.34   67.91
1696193237   775  0.517  0.95  0.49   -0.31   72.12
1699263827  1133  0.898  0.65  0.58   -0.45   57.95
1699268612   854  0.712  0.85  0.60   -0.35   66.71
1699270145   826  0.622  0.90  0.56   -0.31   69.64
1710055644   732  0.498  1.02  0.51   -0.50   84.07
1710059934   692  0.530  1.08  0.57   -0.43   96.03
```

**Appendix B. Estimation of uncertainty**

**B.1 Measurement model.** Let $(l_{ci}, s_{ci})$ denote the measured line and sample coordinates (in pixel units) of control point $c$ in image $i$. We model the measurements as the sum of predicted values and error terms,

$$l_{ci} = L_{ci}(\gamma, \mathcal{P}) + \delta l_{ci},$$

$$s_{ci} = S_{ci}(\gamma, \mathcal{P}) + \delta s_{ci}, \tag{B1}$$

where $(L_{ci}, S_{ci})$ denote the predicted line and sample coordinates as a function of the libration amplitude, $\gamma$, and other parameters, $\mathcal{P}$, including the control-point locations, $\{(X_c, Y_c, Z_c)\}$, and image center coordinates, $\{(x_i, y_i)\}$. We treat the error terms, $\delta l_{ci}$ and $\delta s_{ci}$, as statistically independent, with probability distributions that have zero means and finite variances.

**B.2 Parameter estimation.** We estimate parameters by minimizing the mean squared residual (MSR, with units of pixels squared),

$$R^2(\gamma, \mathcal{P}) = \frac{1}{N} \sum_{i=1}^{N_I} \sum_{c \in C_i} \{[l_{ci} - L_{ci}(\gamma, \mathcal{P})]^2 + [s_{ci} - S_{ci}(\gamma, \mathcal{P})]^2\},$$

(B2)

where $N_I = 340$ is the number of images, $C_i$ denotes the set of control points measured in image $i$ ( Most images had 5−30 control-point measurements; a few dozen had fewer than 5, and several dozen had 30 to ~100 control point measurements. Most control points appeared in at least 10 images.) and $N = 11{,}746$ is the total number of measurements (two for each of 5873 control point/image instances). The minimization is performed in two steps. First, over a grid of $\gamma$ values, we minimize $R^2(\gamma, \mathcal{P})$ with respect to $\mathcal{P}$ using an iterative solver tailored to control-point estimation. This produces points on a $\mathcal{P}$-optimized MSR curve,

$$\tilde{R}^2(\gamma) = R^2(\gamma, \hat{\mathcal{P}}_\gamma),$$

(B3)

where $\hat{\mathcal{P}}_\gamma$ denotes the optimal value of $\mathcal{P}$ for a given value of $\gamma$. We locate the minimum by fitting the points to a cubic polynomial in the vicinity of the smallest $\tilde{R}^2$ on the grid. In the vicinity of the minimum, a parabola alone fits the MSR curve quite well; we keep the cubic term to capture any slight asymmetry in the curve. The minimum of the polynomial defines the best-fit libration amplitude, $\hat{\gamma}$, and its associated MSR, $\tilde{R}^2$; this also implicitly identifies the overall best-fit values for the remaining parameters, $\hat{\mathcal{P}}$. As explained in section B1, three image-center coordinates are held fixed in the optimization, to tie the control-point solution to a previously-used coordinate system. In total, $M = 2139$ parameters are optimized (1464 control-point coordinates, $680 - 2 \times 3 = 674$ image-center coordinates, and the libration amplitude).

We expect the errors to be dominated by pixel-level measurement errors, so that the error variances should have similar scales for all measurements, despite the physical pixel scale varying greatly among the images (the pixel scale ranged from 26 m px$^{-1}$ to 3.1 km px$^{-1}$; relatively few images had scales at the extremes of this range). This motivated our choice to weight the residuals equally in Equation B2. We verified that this assumption is self-consistent by plotting histograms of the residuals for the line and sample measurements in images well-populated with control points, based on the best-fit parameters; visual comparison of the histograms showed no evidence that the error variance differs among the images. We thus based our uncertainty quantification on the assumption of a common error-variance, $\sigma^2$, across all measurements.

Note that in this setting, to the extent that the parameter uncertainties are small enough that linearizing the model about the best-fit parameters is accurate, the Gauss-Markov theorem provides a formal motivation for minimizing the unweighted MSR (it produces the best linear unbiased parameter estimates for a linear model). Alternatively, without making the linearity assumption, but instead assuming independent, equal-variance normal distributions for the noise, the *likelihood function* for $\gamma$ and $\mathcal{P}$ (i.e., the probability for the measurements, as a function of the values of all of the parameters) is a product of normal distributions for each measurement. The logarithm of the likelihood function (up to an unimportant constant), with $\sigma$ known, is then $-NR^2(\gamma,\mathcal{P})/(2\sigma^2)$ (the sum of the exponents in the normal distribution probability density functions), so the minimum MSR parameter estimate is also the maximum likelihood estimate; and either likelihood ratio or Bayesian methods may be used to quantify uncertainties. In addition, $-N\tilde{R}^2(\gamma)/(2\sigma^2)$ is the logarithm of the *profile likelihood* (which is the likelihood as a function of $\gamma$, maximized with respect to $\mathcal{P}$ for each choice of $\gamma$; see Lampton et al. (1976) for use in astronomy, and Barndorf-Nielson and Cox (1994) for a thorough statistical treatment.). The profile likelihood is a widely used tool for eliminating nuisance parameters, i.e., uninteresting parameters needed for modeling the data whose uncertainty must be accounted for in estimating a subset of interesting parameters. For this problem, there are 2138 nuisance parameters, and it is important to account for their uncertainties in summarizing the implications of the data for the one interesting parameter, $\gamma$. We do not adopt the Gaussian noise assumption for our analysis, but we use ideas from profile likelihood analysis to define confidence regions for $\gamma$; see Section B4.

**B.3 Noise-variance estimation.** To quantify the uncertainty in $\gamma$, we need to estimate the noise variance. Let $\gamma_*$ and $\mathcal{P}_*$ denote the true values of the parameters. Were these values known, $R^2(\gamma_*, \mathcal{P}_*)$ would provide an unbiased estimate of $\sigma^2$. Instead, $\gamma_*$ and $\mathcal{P}_*$ are unknown, estimated by minimizing $R^2$, and we expect $\hat{R}^2 = R^2(\hat{\gamma}_*, \hat{\mathcal{P}}_*)$ to *under*estimate $\sigma^2$. To the extent that a linearized model is a good approximation to the true model, standard least-squares theory shows that the expectation value of $\hat{R}^2$ satisfies

$$\langle \hat{R}^2 \rangle = \frac{N-M}{N}\sigma^2 , \tag{B4}$$

indicating that $\hat{\sigma}^2 \equiv N\hat{R}^2/(N-M)$ is an approximately unbiased estimator of $\sigma^2$. Put differently, asymptotically $N\hat{R}^2/\sigma^2$ has an expectation value of $N-M$, corresponding to "minimum $\chi^2$" fitting with $N-M$ degrees of freedom. The familiar $N/(N-1)$ adjustment of the sample variance is an example of this same line of reasoning, for an $M=1$ case. We describe a test of the adequacy of this $\hat{\sigma}$ estimate below; it indicates that the approximate linearity assumption underlying (B4) is justified (which we expect, a posteriori, due to the small scale of the uncertainties).

**B.4 Studentized residual bootstrap resampling.** We want to quantify the statistical uncertainty in the $\gamma$ estimate, using the data-based estimate of the error variance, but not presuming that the errors follow a normal distribution, and not assuming the adequacy of a linearized model. We use bootstrap resampling to calibrate a confidence region for $\gamma$ (Davidson and Hinkley, 1997). In a regression (function-fitting) setting, a recommended semiparametric bootstrap approach is to resample the residuals, i.e., to use the collection of residuals from the best-fit model,

$$\Delta l_{ci} = l_{ci} - L_{ci}(\hat{\gamma}, \hat{\mathcal{P}}),$$

$$\Delta s_{ci} = s_{ci} - S_{ci}(\hat{\gamma}, \hat{\mathcal{P}}), \tag{B5}$$

as a stand-in for the actual distribution of the errors. However, the reasoning of the previous paragraph indicates that the residuals will have a variance that (in expectation) underestimates the actual error variance.

The *Studentized residual bootstrap* accounts for this by scaling up the residuals by a factor $f = \sqrt{N/(N-M)}$ before resampling. "Studentized" refers to a similar scaling that appears in the Student's $t$ statistic. In a linear regression setting, Studentization includes an additional covariate-dependent leverage factor that is not readily available in complex, nonlinear models such as ours. There are alternative residual adjustments in the bootstrap literature (Davidson and Hinkley, 1997; Bickel and Freedman, 1983; Weber, 1984). Residuals should also be *centered* before resampling if their mean differs significantly from zero; here the means of the line and sample residuals were negligible.

To establish the statistical properties of our parameter estimation and confidence region procedures, we use the predictions of the best-fit model as a stand-in for the true model, and

quantify the variability of a procedure by applying it to mock data sets with simulated measurements:

$$l'_{ci} = L_{ci}(\hat{\gamma}, \hat{\mathcal{P}}) + f\Delta l_J,$$

$$s'_{ci} = S_{ci}(\hat{\gamma}, \hat{\mathcal{P}}) + f\Delta s_K, \quad (B6)$$

where the multi-indices $J$ and $K$ are chosen at random (with replacement) from the set of $(c, i)$ indices labeling the residuals. We simulated 3000 data sets, the number needed to accurately calibrate 95.4% ("2-σ") confidence regions (Booth and Sarkar 1998); larger confidence levels require larger bootstrap sample sizes.

We analyze each mock data set employing the same algorithm used for the real data set. The left pair of panels (with shared ordinate) in Fig. B1 depicts the analysis. The left panel shows the $\tilde{R}^2(\gamma)$ curves for a random subset of 50 of the 3000 bootstrapped data sets; the dots indicate the minimum MSR $\hat{\gamma}$ estimates for each case, and the vertical dashed line indicates the "true" value of $\gamma$ underlying the simulations (i.e., the best-fit value found using the actually observed data). Fig. 3 in the main paper shows a histogram of all 3000 minimum MSR estimates. In Fig. B1, the adjacent panel shows the 50 displayed minimum MSR values, $\tilde{R}^2(\hat{\gamma})$, as dots projected along the ordinate; the histogram summarizes the distribution of values from all 3000 data sets. The green dashed line indicates the minimum MSR value found using the real data. The real-data value is typical of what was seen in the simulations. This is a basic test of self-consistency of the bootstrap simulation, indicating that the bootstrap residuals have the same scale as the observed residuals. Studentization of the residuals is essential for this self-consistency; a calculation using unscaled residuals in the bootstrap produces mock data sets with minimum MSR values very different from the minimum MSR found with the actual data.

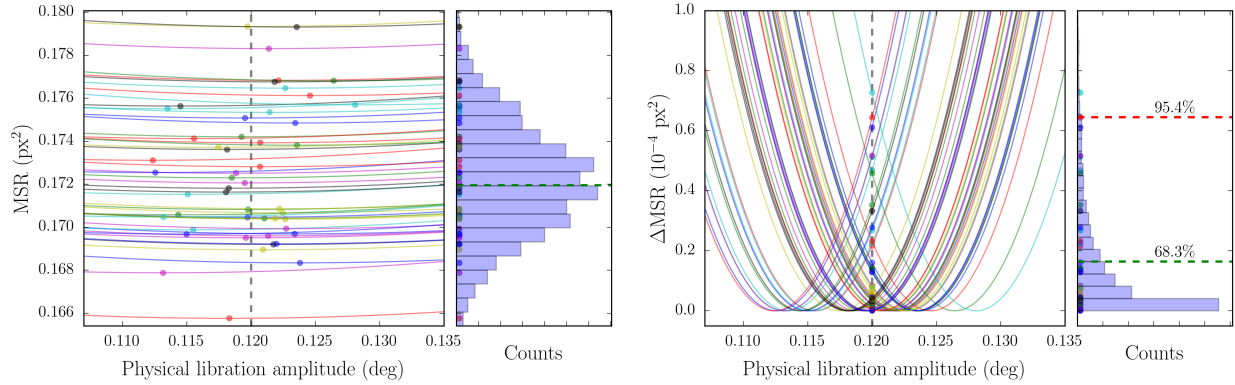

**Figure B1.** Ingredients for bootstrap confidence interval calibration. *Left set:* MSR vs. $\gamma$ curves for 50 (of 3000) simulated data sets created by adding resampled scaled residuals to the predictions of the model that best fits the observed data. Dots indicate the best-fit libration amplitude, $\hat{\gamma}_b$, for each data set; the dashed vertical line indicates the best-fit value for the observed data, $\hat{\gamma}$, which plays the role of the "true" value in the simulations. Side panel shows the MSR for the 50 $\hat{\gamma}_b$ values as dots along the ordinate, and a histogram of all 3000 values. The dashed green line indicates the MSR value for the observed data. *Right set:* Change in MSR with respect to its minimum value for the same 50 data sets; dots indicate $\Delta R^2$ for the true value, $\hat{\gamma}$, i.e., how far up the MSR curve one must go to reach the true parameter value. Side panel shows these $\Delta R^2$ values as dots along the ordinate, and a histogram of all 3000 values. Green and red lines show $\Delta R^2$ values containing 68.3% and 95.4% of the simulated cases, respectively.

**B.5 Confidence regions.** For a univariate linear model with normally distributed data, the MSR curve is a parabola, and a conventional confidence region with confidence level $C$ may be defined as the set of parameter values with MSR within a constant offset, $\Delta R_C^2$, from the bottom of the parabola (in two dimensions, the analogous rule corresponds to drawing a contour of the $R^2$ function). For example, for a "1-σ" interval with $C \approx 0.683$, one uses $N\Delta R_C^2/\sigma^2 = 1$, and for a "2-σ" interval with $C \approx 0.954$, one uses $N\Delta R_C^2/\sigma^2 = 4$ (i.e., with 4 being the squared distance from the best-fit value to the 2-sigma boundary). We define a confidence region for $\gamma$ analogously, relying on the bootstrap samples to identify appropriate values for the offset. The rightmost set of panels in Fig. B1 illustrates the procedure we follow to find values of $\Delta R_C^2$. The larger panel shows the same 50 MSR curves as discussed previously, but with the minimum MSR values subtracted, displaying

$$\Delta R_b^2(\gamma) = R_b^2(\gamma) - R_b^2(\hat{\gamma}_b) , \qquad (B7)$$

where the $b$ subscript indicates a quantity associated with a particular bootstrap sample.

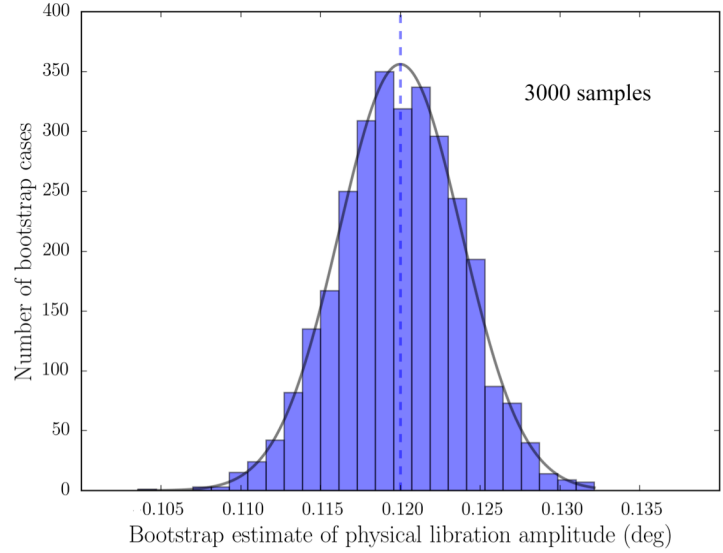

**Figure B2.** Histogram of best-fit parameter values, $\hat{\gamma}_b$, for 3000 simulated data sets. Dashed vertical line indicates the best-fit value for the observed data. A percentile confidence region with approximate confidence level $C$ may be found by excluding a fraction of samples $(1 - C)/2$ from each side of the distribution. In contrast to the procedure of Fig. B1, the percentile procedure uses only the locations of the minima of the MSR curves, ignoring variability in the shapes of the curves.

The vertical dashed line again indicates the "true" value of $\gamma$ underlying the simulations, i.e., $\hat{\gamma} = 0.120°$. The dots indicate the $\Delta R_b^2$ values for $\hat{\gamma}$ showing how far up the curve one must go to just reach the "true" value underlying the simulations. The adjacent panel shows these $\Delta R_b^2(\hat{\gamma}_b)$ values as dots projected along the ordinate, with a histogram calculated using all 3000 bootstrap data sets. The critical values defining (approximate) 1-σ and 2-σ confidence regions are just the values bounding (from above) the appropriate percentage of $\Delta R_b^2(\hat{\gamma}_b)$ values. Here the critical values satisfy $N\Delta R_C^2/\sigma^2 \approx 0.92$ and 3.6 for 68.3% and 95.4% confidence levels (compared to 1 and 4 in the univariate linear, normal case). The 95.4% confidence level is ± 0.0078°.

A simpler, more approximate confidence region may be found using a *percentile bootstrap confidence region*. This procedure ignores variability in the shapes of the MSR curves, focusing only on the distribution of the locations of the best-fit points. Fig. B2 shows a histogram of the bootstrap best-fit values, $\hat{\gamma}_b$ (This is a slight variation on Fig. 1d in the main text.). The mean of the $\hat{\gamma}_b$ values is equal to the "true" value to within Monte Carlo uncertainties, indicating that the minimum-MSR estimator is approximately unbiased. A percentile confidence region with approximate confidence level $C$ may be found by excluding a fraction, $(1 - C)/2$, of samples from each side of the distribution. For this procedure to give approximately correct regions, the distribution of estimates should be nearly normal (Gaussian) with negligible bias, which is the case here. In particular, the distribution must be symmetric. If it were, for instance, skewed to the right, that would indicate a tendency to overestimate the true value of the parameters, so a sound confidence region should be skewed to the left. Yet in such a case, the percentile region would itself be skewed to the right, in exactly the opposite manner needed to accurately convey the effect of the errors. For our data and model, symmetry and approximate normality hold, and the percentile and $\Delta R_C^2$ regions agree to within Monte Carlo uncertainties.